# Adaptive Quantum Tomography in a Weak Measurement System with Superconducting Circuits

Hyeok Hwang[1], JaeKyung Choi[1], and Eunseong Kim[1]*

[1] *Department of Physics, Korea Advanced Institute of Science and Technology (KAIST), Republic of Korea*

**Abstract**

Adaptive tomography has been widely investigated to achieve faster state tomography processing of quantum systems. Infidelity of the nearly pure states in a quantum information process generally scales as $O(1/\sqrt{N})$, which requires a large number of statistical ensembles in comparison to the infidelity scaling of $O(1/N)$ for mixed states. One previous report optimized the measurement basis in a photonic qubit system, whose state tomography uses projective measurements, to obtain an infidelity scaling of $O(1/N)$. However, this dramatic improvement cannot be applied to weak-value-based measurement systems in which two quantum states cannot be distinguished with perfect measurement fidelity. We introduce in this work a new optimal measurement basis to achieve fast adaptive quantum state tomography and a minimum magnitude of infidelity in a weak measurement system. We expect that the adaptive quantum state tomography protocol can lead to a reduction in the number of required measurements of approximately 33.74% via simulation without changing the $O(1/\sqrt{N})$ scaling. Experimentally, we find a 14.81% measurement number reduction in a superconducting circuit system.

## Introduction

Quantum state tomography (QST) is a procedure of finding an adequate description of a quantum state based on tomographic data collected from a series of measurements [1–17]. For example, a goal of QST may be to identify an estimator ($\hat{\rho}$) that describes the state of a qubit ($\rho$). Various approaches to optimize QST have been studied both in the direction of finding reasonable estimators from given tomographic data and in the direction of modifying the sampling strategies applied to obtain the tomographic data.

Several different approaches to the identification of a quantum state from tomographic data have been investigated for decades. One example is direct inversion tomography that designates the most probable state as the estimator for given tomographic data [5]. However, the resulting estimator from this method can be an unphysical state (outside the Bloch sphere) due to statistical uncertainty. Another example approach is maximum likelihood estimation that refines the probabilistic interpretation of direct inversion tomography by limiting the possible range of qubit state vectors to the inside of the Bloch sphere; in this case, though, the maximum of the likelihood function may not be unique, which is known as rank deficiency [2,5,8,12,13,15,16,18–20]. To eliminate this rank-deficient property of the likelihood function, a third approach to the identification of a quantum state from tomographic data is Bayesian mean estimation [5,9,18,20–22], which interprets the likelihood function as a weighting factor for a given density operator. As the resulting estimator from Bayesian mean estimation is never rank deficient, this approach is widely used for QST.

In terms of sampling strategies for tomographic data, in addition to using Cartesian axes, the use of four axes passing through the vertices of a tetrahedron has been investigated [9,23], for example. But because estimators converge at different rates depending on the quantum state [5,13,18,19,24], a fixed set of sampling axes is not optimal for every quantum state simultaneously. To overcome this limitation, strategies to construct optimal measurement axes from prior information of quantum states have been suggested, which are known as adaptive quantum state tomography (AQST) [18,19,21–26].

Theoretical methodologies for AQST have focused on optimizing measurement configurations, utilizing mutually unbiased observables [13,27], and expanding the number of adaption steps [24]. In addition, experimental realizations of AQST have been reported in projective measurement systems [18,19,21,22,24]. Despite this progress, one area that is relatively less explored is the experimental realization of AQST for weak measurement systems, which are systems that acquire weak values of observables through state-entangled classical fields [28–31]. Considering the recent spotlight on superconducting circuit systems [31–38], a type of weak measurement system, we suggest in this work a methodology for AQST in a transmon qubit system, which is one of the most widely investigated segments of superconducting circuit systems.

Several kinds of metrics, such as infidelity [5,12,13,18,19,24], can quantify the error of estimation in QST. Because these metrics are related to the power of estimation error, the proposed AQST approach employs a variance of estimation error that is dependent on the state of a qubit to minimize the variance of estimation error. The process involves two stages: a relatively small number of ensembles are initially measured to provide a rough estimation of the state of a qubit, and then a large number of ensembles are measured to obtain a precise estimation after applying a feedback drive to minimize variance. The reduction in the number of required measurements to achieve a specific error bound is related to the change of this variance. Hence, we depict the effectiveness of our AQST by comparing the relation between the number of measurements versus the variance of the measured data of our AQST case to that of a standard QST case with both simulation and experiment. We find a remarkable reduction in the number of required measurements to get a specific error, and thus expect that the new AQST process can offer advantages in state readout.

## Result

Several metrics can be used to quantify the difference between the qubit state $\rho$ and the estimator $\hat{\rho}$ from QST. Example metrics include the Hilbert–Schmidt distance $d_{HS}(\rho, \hat{\rho}) = \text{Tr}(\rho - \hat{\rho})^2$ that is related to the probability of getting estimator $\hat{\rho}$ from given qubit state $\sigma$, trace distance $d_{tr}(\rho, \hat{\rho}) = \text{Tr}|\rho - \hat{\rho}|$ that is related to single shot error, and infidelity $1 - F(\rho, \hat{\rho}) = 1 - \text{Tr}\left[\sqrt{\sqrt{\rho}\hat{\rho}\sqrt{\rho}}\right]^2$ that is frequently used for not only its sensitivity but also its physical meaning for pure state cases [5,12,13,18,19,24]. These metrics are proportional to the power of the estimation error $\Delta = |\rho - \hat{\rho}|$ [12,18], so minimizing the variance of $\Delta$ is a main concern of QST. Therefore, we studied the statistics of the variance of $\Delta$ as follows.

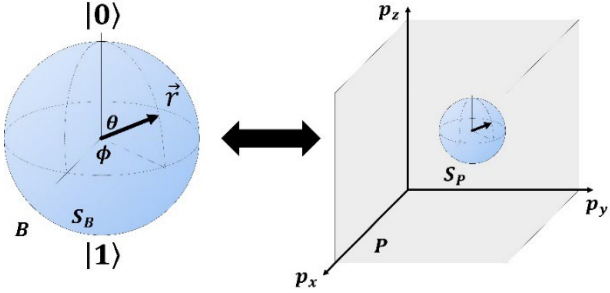

FIG. 1: The quantum state of a qubit can be described as a point $\vec{r}$ on sphere $S_B$ in Bloch space B (left). By mapping $S_B$ to cubic probability space P, point $\vec{p} = \sum_{k=x,y,z} p_k \hat{k}$ can represent quantum state tomography (QST) data along 3 axes (right), where $p_k$ is defined as the probability of measuring 0 along the k-axis (k = x, y, z). The states of the Bloch sphere $S_B \subset B$ have 1:1 correspondence with the sphere $S_P \subset P$.

To find the relation between estimation error $\Delta$ and tomographic data, the relation between point $\vec{p}$ in probability space P and point $\vec{r}$ in Bloch space B needs to be checked. The probability of getting a measurement outcome corresponding to 0 or 1 for qubit state $|\psi\rangle = |g\rangle$ or $|e\rangle$ is denoted as $P_{|i\rangle \to j}$ ($i \in \{g, e\}, j \in \{0, 1\}$). Whether the measurement outcome belongs to 0 or 1 depends on the classification criteria between 0 and 1, and thus the probability $P_{|i\rangle \to j}$ also depends on the classification criteria. We denoted the probability of measuring 0 along the k-axis (k = x, y, z) as

$$p_k = P_{|g\rangle \to 0}\langle g|\psi\rangle + P_{|e\rangle \to 0}\langle e|\psi\rangle,$$

and then the vector $\sum_{k=x,y,z} p_k \hat{k}$ resides in cubic probability space P = $\{(x, y, z) | 0 \leq x, y, z \leq 1\}$ (Fig. 1).

A quantum state of a qubit

$$|\psi\rangle = \cos\theta|g\rangle + e^{i\phi}\sin\theta|e\rangle$$

can be represented as the point $\vec{r} = (r_x, r_y, r_z)$ in Bloch space. We represent the relation between $r_k$ and $p_k$ as $p_k = (\alpha_0 + \alpha_1 r_k)/2$, where $\alpha_{0,1} = P_{|g\rangle \to 0} \pm P_{|e\rangle \to 0}$ ($P_{|e\rangle \to 0} \leq p_k \leq P_{|g\rangle \to 0}$). Because tomographic data follows binomial statistics $X_k \sim B(n, p_k)$ along each k-axis, for an estimator represented as $\hat{p} = (\hat{p}_x, \hat{p}_y, \hat{p}_z)$ with $\hat{p}_k = X_k/(N/3)$, the mean and standard deviation of $\hat{p}$ can be found as follows:

$$X_k \sim B(N/3, p_k) \to m(X_k) = (N/3)p_k, \sigma(X_k)$$
$$= \sqrt{(N/3)p_k(1 - p_k)} \to m(\hat{p}_k)$$
$$= p_k, \sigma(\hat{p}_k) = \sqrt{\frac{p_k(1 - p_k)}{(N/3)}}. (1)$$

According to the above representation of standard deviation, $\sigma(\hat{p}_k)$ scales as $O(N^{-1/2})$ except for extreme cases when $p_k$ or $(1 - p_k)$ scales as $O(N^{-1})$ [19]. For a weak measurement system, as the measurement time increases, the qubit experiences more energy relaxation. On the other hand, as the measurement time decreases, the fidelity of the measurement also decreases due to becoming quantum non-demolition measurements. Therefore, the probability $p_k$ is always less than 1, so the standard deviation $\sigma(\hat{p}_k)$ scales as $O(N^{-1/2})$ for weak measurement systems.

One objective of our AQST methodology is to minimize the above standard deviation of $\Delta = |\hat{r} - \vec{r}|$, where

$$\Delta = \left|\sum_k \hat{k}(\hat{r}_k - r_k)\right| = \frac{2}{\alpha_1}\left|\sum_k \hat{k}(\hat{p}_k - p_k)\right|. (2)$$

Every measurement is an independent event, so the variance of the estimation error is as below:

$$\sigma^2(\Delta) = \sigma^2\left(\frac{2}{\alpha_1}\left|\sum_k \hat{k}(\hat{p}_k - p_k)\right|\right) = \frac{4}{\alpha_1^2}\sum_k \left(\frac{3}{N}\right)p_k(1 - p_k)$$
$$= \frac{12}{N\alpha_1^2}\left[\frac{3}{4} - \sum_k \left(p_k - \frac{1}{2}\right)^2\right]$$
$$= \frac{12}{N\alpha_1^2}\left(\frac{3}{4} - |\vec{R}|^2\right). (3)$$

Therefore, the variance $\sigma^2(\Delta)$ is represented as a function of $\vec{R} \equiv \sum_{k=x,y,z} \hat{k}(p_k - 1/2)$, which is the difference between estimator $\vec{p} \in$ P and $P_O(1/2, 1/2, 1/2)$, by the orthogonality of our measurement axes.

By Eq. (3), the minimum variance of $\Delta$ can be achieved by maximizing the difference $|\vec{R}|$. The relation between the qubit state and the minimum variance of $\Delta$ is depicted in Fig. 2. The qubit estimator for the minimum and maximum variance is depicted as $V_{min}$ (red) and $V_{max}$ (blue), respectively. If we have information about the qubit state from a small number of QST data, by providing feedback that forces the qubit to be in the state corresponding to $V_{min}$, the variance of the QST can be minimized.

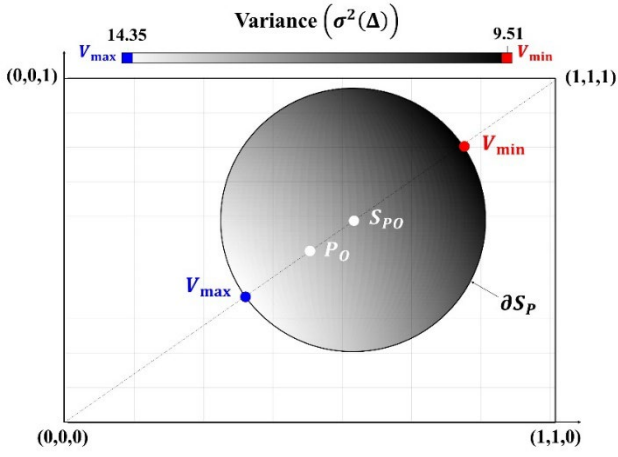

FIG. 2: Variance color map of qubit states on $S_P$ with $\alpha_0 =$ 1.178, $\alpha_0 = 0.764$ depicted on the probability space P cut by four vertices denoted in figure. The centers of P and $S_P$ are denoted as $P_O$ and $S_{PO}$, respectively. According to Eq. (2), the variance of the estimator has a minimum and maximum value at $V_{min}$ and $V_{max}$, which are the farthest and nearest points on $S_P$ from point $P_O$, respectively. The four mentioned points are on the line connecting (0, 0, 0) and (1, 1, 1).

Based on the above discussion, we suggest a new AQST process consisting of two steps, namely pre- and post-estimation, that reduces the standard deviation of tomographic data by providing adaptive feedback in a weak measurement apparatus. Schematics of the proposed method and our experimental setup are illustrated in Fig. 3. With the new AQST method, the $O(N^{-1/2})$ scaling of the standard deviation of the data as in standard QST cases is conserved, but the magnitude of the standard deviation is minimized through geometrical optimization in Bloch space.

We use a three-dimensional (3D) transmon qubit as implemented in Fig. 3b. The transmon qubit, with a transition frequency of 5.7959 GHz, is mounted in a 5.0714 GHz Al superconducting cavity with a coupling strength of $g/2\pi$ = 23.90 MHz and a Stark shift of $\chi/2\pi$ = 0.79 MHz. The qubit relaxation time T1 is 1.7604 μs and decoherence time T2 is 1.3503 μs. The linewidth of the cavity $\kappa/2\pi$ is 0.41 MHz. We conduct each qubit state readout with microwave pulses in a heterodyne detection scheme (Fig. 3). The length of the control pulse is 180 ns for each rotation, and the length of the measurement pulse is 2 μs long, which gives the maximum distinguishability of $|g\rangle$ and $|e\rangle$ in the IQ plane.

We prepare an ensemble of qubits in $|\psi\rangle_{worst}$, which is a state corresponding to $V_{max}$, and implement AQST by providing feedback to $|\psi\rangle_{best}$, which is a state corresponding to $V_{min}$. The parameters of our system are summarized in Table 1. The states $|\psi\rangle_{worst}$ and $|\psi\rangle_{best}$ can be calculated as follows:

$$|\psi\rangle_{worst} = \begin{bmatrix} \sqrt{\dfrac{3-\sqrt{3}}{2}} \\ e^{i\frac{5\pi}{4}}\sqrt{\dfrac{3+\sqrt{3}}{2}} \end{bmatrix}, |\psi\rangle_{best} = \begin{bmatrix} \sqrt{\dfrac{3+\sqrt{3}}{2}} \\ e^{i\frac{\pi}{4}}\sqrt{\dfrac{3-\sqrt{3}}{2}} \end{bmatrix}. \quad (4)$$

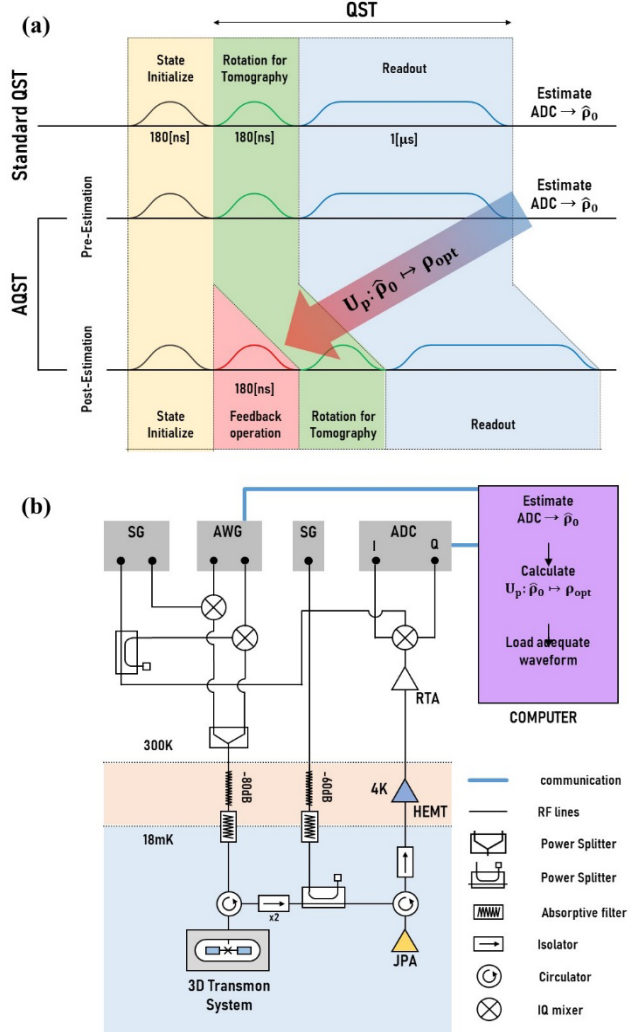

FIG. 3: Overview of the proposed AQST method. (a) Comparison of the standard QST process and proposed AQST process consisting of two steps. Pre-estimation is first implemented where a relatively small number of measurements are collected by an analog-to-digital converter (ADC) and used to find the first estimator $\hat{\rho}_0$. We calculate the adequate feedback function $U_P$ to transform $\hat{\rho}_0$ to $\rho_{opt}$. Post-estimation is then implemented via the calculated feedback operation with a relatively large number of measurements. The result is a reduced estimation error from the same number of measurements. (b) Schematic of our experimental setup. A three-dimensional transmon qubit is mounted in an Al cavity on a mixing chamber stage at 18 mK in a dilution refrigerator. The output signal is amplified by three amplifiers: a Josephson parametric amplifier (JPA), high electron mobility transistor (HEMT),

and room temperature amplifier (RTA) at 18 mK, 4 K, and 300 K, respectively. The feedback operation illurrated in (a) is calculated from the data collected by pre-estimation though a computer, and the resulting pulse sequence corresponding to the calculated feedback operation is loaded to an arbitrary waveform generator (AWG).

TABLE I: Parameters of our transmon system (up to significant figures)

| Variable | $\omega_c/2\pi$ [GHz] | $\omega_q/2\pi$ [GHz] | $T_1$ [μs] | $T_2$ [μs] |
|---|---|---|---|---|
| Value | 5.0714 | 5.7959 | 1.7604 | 1.3503 |
| Variable | $\chi/2\pi$ [MHz] | $g/2\pi$ [MHz] | $\kappa/2\pi$ [MHz] | |
| Value | $7.9\times10^{-1}$ | $2.309\times10^{1}$ | $4.1\times10^{-1}$ | |

Information about the measured photons corresponds to a point on the IQ plane. After the qubit is initialized into $|g\rangle$ or $|e\rangle$, two Gaussian envelopes respectively representing $|g\rangle$ and $|e\rangle$ are measured and analyzed by fitting with Python. Then we normalize the population along the normalized coordinate $\lambda_D$, as represented in Fig. 4a.

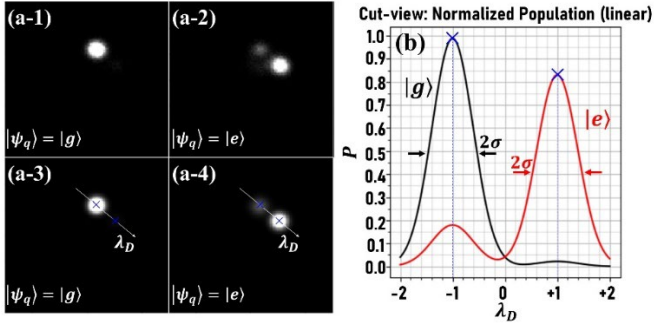

FIG. 4: State-dependent cavity photon responses in the IQ plane and corresponding analysis. (a-1) and (a-2) depict the populations of readout photons in the IQ plane for the ground and excited qubit state, respectively. (a-3) and (a-4) fit the respective populations with two Gaussian envelopes. We normalize the distance between the two envelopes along the peak-to-peak direction with normalized coordinate $\lambda_D$. (b) Plot of normalized populations of readout photons for $|g\rangle$ (black) and $|e\rangle$ (red). The standard deviation for each envelope is $2\sigma = 0.7896$ after normalization. IQ information for each readout photon is converted to a normalized coordinate, and we determine that the state is 0 if $\lambda_D < \lambda_c$ for an adequate criteria $\lambda_c$.

The selection of the criteria to discriminate between 0 and 1 influences the tomographic results. In systems based on weak measurement that require a finite measurement time, the relaxation, decoherence, and thermal excitation of a qubit provide an inevitable asymmetry of the $|g\rangle$ and $|e\rangle$ populations in the IQ plane. In particular, because the coherence time of a qubit is not infinite, a transmon in a cryogenic environment is relatively more affected by relaxation and decoherence than by thermal excitation, so the measured photons are more likely to be found in the $|g\rangle$ envelope.

The population skewness is parameterized by $\alpha_{0,1}$ as a function of the selected criteria $\lambda_c$, as shown in Fig. 5a.

For the given system, the variance of measured data per measurement number N follows Eq. (3), so the minimization condition of criteria $\lambda_c$ is dependent on $|\psi\rangle_q \in S_B$. In the present transmon system, optimization of the variance for $|\psi\rangle_q = |\psi\rangle_{best}$ or $|\psi\rangle_q = |\psi\rangle_{worst}$ is carried out at $\lambda_c = 0.11$, which is the general criteria typically used for state readout because it gives the maximum speed of convergence.

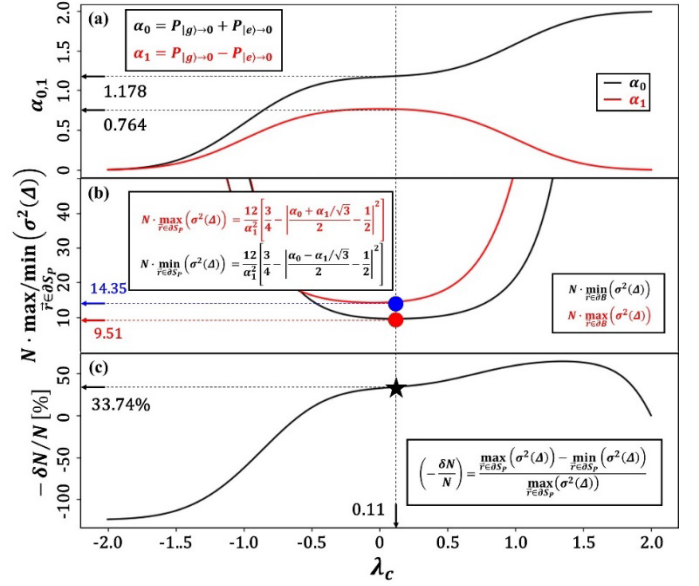

FIG. 5: Calculation process to get a reduced number of required measurements as the criteria $\lambda_c$ changes. The variables $\alpha_{0,1}$, the extremums of variance, and the reduced required number of measurements to get a specific error bound as a function of criteria $\lambda_c$ are shown. The optimal $\lambda_c$ that minimizes the minimum variance is 0.11, which is represented as the vertical dashed line. The equations that represent the meanings of the related y-axis are given in each panel. (a) The $\alpha_1$ variable represents the extent that the two qubit states are decomposed. The $\alpha_1$ maximum around $\lambda_c = 0.11$ represents that the two states are well discriminated. (b) Plot of the extremums of variance per data. A crossing is seen at $\lambda = -0.5828$ where the point $S_{PO}$ crosses the point $P_O$ such that the farthest and nearest points are exchanged. (c) A reduction in the required number of measurement data of around 33.74% at $\lambda_c = 0.11$ is found. The number of measurement data required to get an equal error bound is proportional to variance $\sigma^2$ because the statistics for the measurement data follow binomial distribution normalized by N.

At this criteria, the variance per measurement [N × $\sigma^2(\Delta)$] for $V_{min}$ is 33.74%, which is smaller than that for $V_{max}$. The required number of measured data to get the same error bound is proportional to the variance of the data because QST data follow binomial statistics, having a variance that is inversely proportional to N. Therefore, the reduction in the required number of measurements ($-\delta N/N$) is expected to be 33.74% if we drive the qubit from $V_{max}$ to $V_{min}$ via feedback. We verified the relative reduction in variance, i.e., the relative reduction in the required number of measurements, by comparing the experimental results of standard QST to those of our proposed AQST, as shown in Fig. 6. Because the variance scales as O(1/N) for both cases, the depicted data in the figure are log-normalized for a clearer description.

The blue data in Fig. 6 reflect the cost of measurement number $N_{pre}$ for pre-estimation, but they cannot reflect the convergence of post-estimation independently. Therefore, additional data points with a subtraction of $N_{pre}$ from the number of data N are plotted in red. The experimental data are plotted with crossed square symbols for both standard QST and AQST schemes. The two dashed horizontal lines are constant y lines fitted to the data of both schemes. From the difference between the two dashed lines, we can see a 14.81% reduction in variance for AQST relative to standard QST. The theoretically expected reduction is 33.74%, which is the difference between the upper bound (corresponding to $V_{max}$) and the lower bound (corresponding to $V_{min}$) of the variance, depicted as solid lines in Fig. 6.

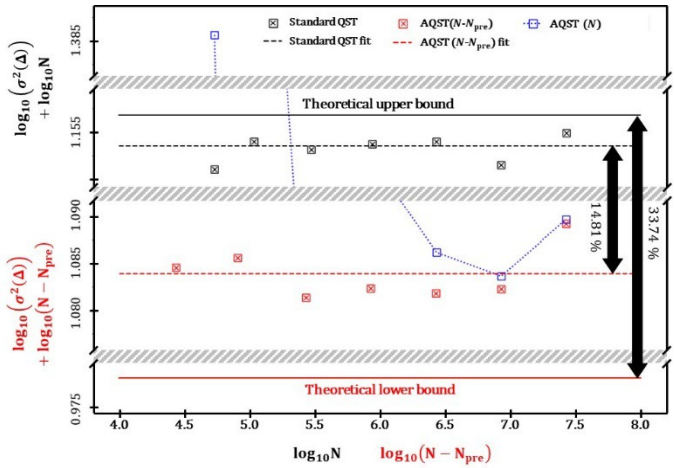

FIG. 6: Convergence of the variance acquired by experiment (crossed square symbols) for standard QST (black) and proposed AQST (red). The graph of variance is normalized by logN. We used $N_{pre}$ = 27000 samples for pre-estimation, so the data points for AQST are shifted to reflect the computational resources for adaptive feedback (blue). Because $N_{pre}$ data for pre-estimation is not reflected in the total variance, the blue line does not scale as O(1/N). To compare the variance of the data from our AQST experiment to the model equation, $N_{pre}$ samples are neglected for a pure comparison (red). For both cases of standard QST and AQST, the variances are well fitted along a constant y line, meaning that they converge with 1/N. The variance decreased by 14.81% for AQST relative to standard QST for both experiment and model. The theoretically expected reduction in variance between the upper and lower bounds for the variance is 33.74%. This theoretical reduction can be accomplished when using a qubit with a long coherence time.

This difference between expected and experimental results may derive from two particular characteristics of our experimental setup. The first concerns the small ratio between $T_2$ and the time for gate operation. The decoherence time ($T_2$ = 1.3503 μs) of our transmon is not long enough to neglect the effect of decoherence during gate operation (180 ns). After providing feedback, we found that the purity of the qubit state decreased significantly to 0.70, even though the direction of the estimator $\hat{r}$ was well aligned with the optimal point. The second reason stems from the relatively small number of measurements we carried out for pre-estimation. As the pre-estimation step involves a much smaller number of data than the post-estimation step, the estimator from the first step was not perfect. Indeed, we found that the estimator from pre-estimation had errors of around 5.9° in the elevation angle and 3.2° in the azimuth angle of the Bloch sphere.

## Discussion

Many researches try to overcome the state-dependent convergence of estimation error in QST by employing the feedback process of AQST. In the same context, the methodology of AQST proposed in this paper has the objective to reduce the error of estimation for the case of weak measurement system. A theoretical investigation of the variance of estimation error gives a minimum variance point $V_{min}$ in a probability space P that is mapped to a state of a qubit $|\psi\rangle_{best}$. The simulated reduction in the variance of estimation error by the proposed AQST is 33.74%, which is based on state-dependent cavity photon responses in the IQ plane of Fig. 4. Although the 14.81% reduction observed in our experiment is smaller than that of simulation, the experimentally achieved reduction can be said to be a significant value.

Since enhancing the precision of QST with a finite number of ensembles is an important research topic, the reduction observed from our AQST process can be a meaningful tool for tomography in weak measurement systems, such as superconducting circuits including transmons. As a continuation of this research, by utilizing qubits with various lifetimes and optimizing the measurement pulses, we can get a broader understanding as well as improve the efficiency of our AQST process.